\documentclass[twocolumn,
               secnumarabic,
               amssymb,
               nobibnotes,
               aps,
               prm,
               superscriptaddress
               ]{revtex4-1}
\usepackage{bm}
\usepackage{graphicx}
\graphicspath{{images/}}

\begin{document}

\title{Strain Fields in Repulsive Colloidal Crystals}%

\author{Bryan VanSaders}%
\affiliation{Department of Materials Science and Engineering, University of Michigan, Ann Arbor, Michigan 48109, United States}

\author{Julia Dshemuchadse}%
\affiliation{Department of Chemical Engineering, University of Michigan, Ann Arbor, Michigan 48109, United States}

\author{Sharon C.\ Glotzer}%
\affiliation{Department of Materials Science and Engineering, University of Michigan, Ann Arbor, Michigan 48109, United States}
\affiliation{Department of Chemical Engineering, University of Michigan, Ann Arbor, Michigan 48109, United States}
\affiliation{Biointerfaces Institute, University of Michigan, Ann Arbor, Michigan 48109, United States}

\date{\today}%

\begin{abstract}
    The concept of a local linear elastic strain field is commonly used in the metallurgical research community to approximate the collective effect of atomic displacements around crystalline defects.
    Here we show that the elastic strain field approximation is a useful tool in colloidal systems.
    For colloidal crystals with repulsive particle interaction potentials, given similar mechanical properties, sharper potentials lead to: 1) free energies of deformation dominated by entropy, 2) lower variance in strain field fluctuations, 3) increased tension-compression asymmetry near dislocation core regions, and 4) smaller windows of applicability of the linear elastic approximation.
    We show that the window of linear behavior for entropic colloidal crystals is broadened for pressures at which the inter-particle separation sufficiently exceeds the range of steep repulsive interactions.
\end{abstract}

\maketitle

\section{Introduction}

Our ability to synthesize nanoparticles with desired shape or surface functionalization continues to improve \cite{Glotzer2007,Nykypanchuk2008,Sacanna2011}, and research on colloidal crystals has largely focused on finding connections interaction potentials and assembled crystal structures \cite{HajiAkbari2009,Wang2015,Ducrot2017,Chen2011,Damasceno2012,Zhang2013,Nykypanchuk2008,vanAnders2015}
As colloidal research continues to develop, it is expected that desired material properties beyond crystal structure will become targets of investigation for colloidal engineering \cite{Kumar2017}.
One such property of interest is the mechanical response of colloidal materials.
Metallurgists have accumulated a wealth of knowledge and models to describe, predict, and explain the various ways in which bulk properties and defects contribute to the mechanical behavior of a material.
A central concept is the linear elastic strain field, which has been used to understand dislocation interaction energies, enabling the creation of high-strength metals.

The time is now ripe to investigate how such concepts can be applied in colloidal matter.
There are two compelling reasons to do so.
The first is to inform the use of colloidal crystals as analogues of atomic crystals.
Given the large size and long time scales of colloidal particles, it is possible to directly observe processes that range from challenging to impossible to observe with atoms.
Therefore it is attractive to re-create the circumstances that surround a hidden atomic process in a colloidal crystal and observe the system's evolution.
Some pioneering work has already been done along this line.
Schall \textit{et al.}~\cite{Schall2004} devised an interferometry method for viewing dislocations in hard-sphere colloidal crystals.
Later, the same authors employed this technique to observe the formation of dislocation loops in colloidal crystals following indentation by sewing needles \cite{Schall2006}.
More recently, Lin \textit{et al.}~\cite{Lin2016} reported a method for directly measuring local strain fields around dislocations and other defects within colloidal crystals using confocal microscopy.
Van der Meer \textit{et al.}~\cite{vdMeer2017} investigated the strain fields that surround point defects in simulated hard-sphere crystals and used them to explain interstitial clustering behavior.
While these works are foundational for establishing the connection between colloidal and atomic material defect dynamics, many unanswered questions remain.
Importantly, it is not yet clear which atomic phenomena can be meaningfully studied \textit{via} colloidal analogues.
By investigating the use of core metallurgical concepts such as linear elastic strain fields in colloids, we aim to further inform the use of these model systems.

The second motivation for our study is to explore what is possible \textit{beyond} phenomena seen in atomic materials.
Colloidal crystals are famous for their exotic optical properties \cite{Yablonovitch1987,Yablonovitch1991,Maldovan2004,Ho1990}.
The creation of metamaterials with exotic mechanical properties requires greater understanding of the mechanical behavior of colloidal assemblies, which is currently lacking \cite{Kim2016,Tretiakov2014,Zaccarelli2001,Zaccone2009}.
Beyond the realization of unusual material properties, there is also compelling evidence that metallurgical concepts are important throughout the assembly processes.
For example, it has been shown that local material strain plays an important role in the evolution of self-limiting assemblies \cite{Grason2016}.
A more complete understanding of the nature of strain fields in colloidal materials enables greater sophistication both in the search for exotic material properties and the exploitation of geometric constraints for assembly engineering \cite{Irvine2010, Meng2014,Azadi2014,Yao2017}.

\begin{figure}[tp]
\centering
\includegraphics[width=0.5\textwidth,keepaspectratio]{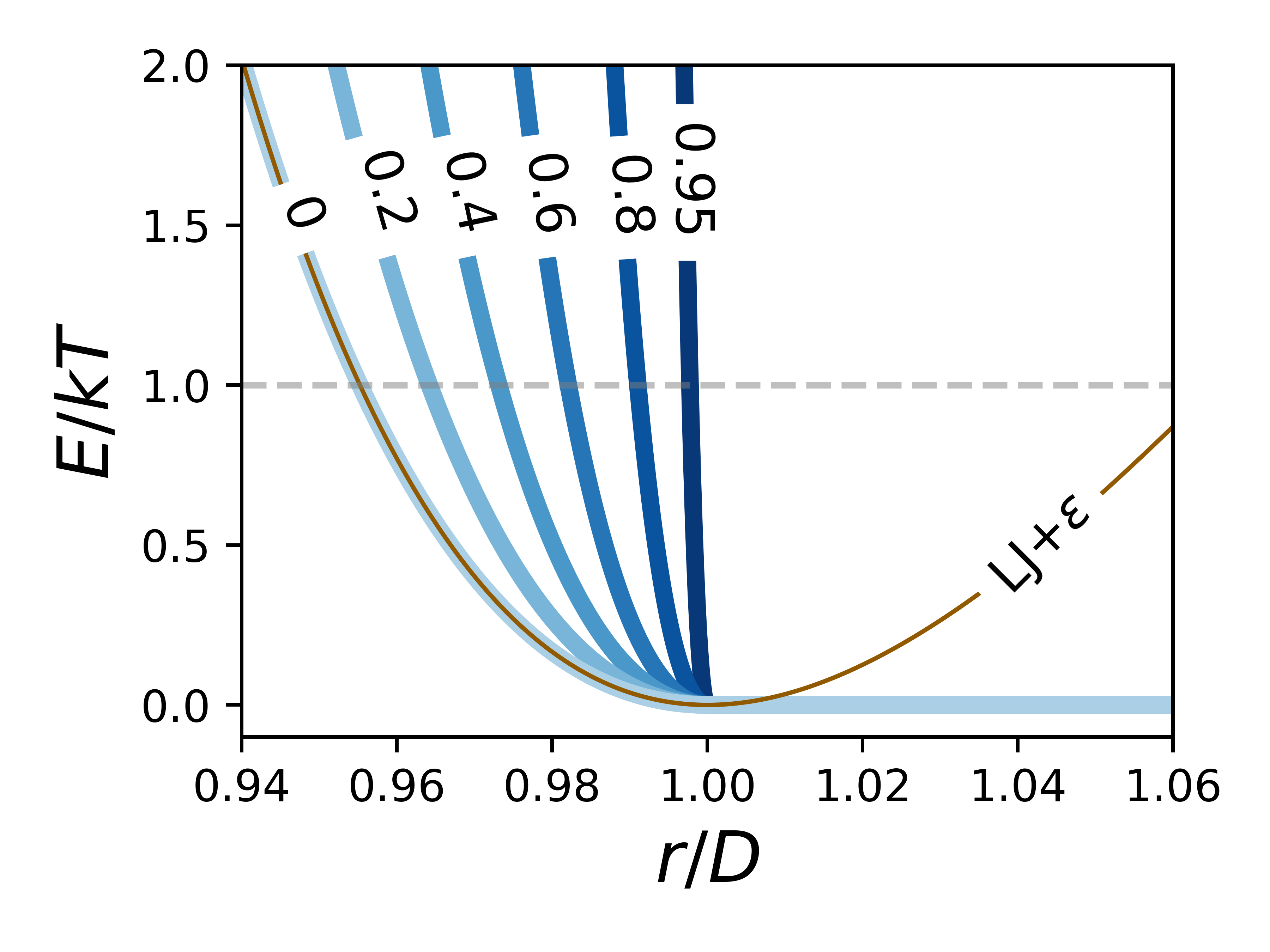}
\caption{
The isotropic pair potentials used in this study.
Darker blue curves are potentials with larger hardness values (labeled on curves).
The Lennard-Jones potential (shifted in energy by $\epsilon$) is also shown for comparison.
}
\label{fig:potentials}
\end{figure}

In this study we focus on a family of isotropic pair potentials (see Fig.~\ref{fig:potentials}) that smoothly range from the canonical Lennard-Jones (LJ) potential used to model atomic systems \cite{LJ1924}:

\begin{eqnarray} \label{eq:LJ}
V_{\mathrm{LJ}}(r)  = & 4 \varepsilon \left[ \left( \frac{\sigma}{r} \right)^{12} - \left( \frac{\sigma}{r} \right)^{6} \right] \quad& r < r_{\mathrm{cut}} \\* \nonumber
                     = & 0 & r \ge r_{\mathrm{cut}}
\end{eqnarray}

\noindent to a hard-sphere potential used to model excluded volume interactions in colloids.
This is done with the Shifted Weeks-Chandler-Andersen (SWCA) \cite{WCA1971} potential:

\scriptsize
\begin{eqnarray} \label{eq:SWCA}
V_{\mathrm{SWCA}}(r)  = & 4 \varepsilon \left[ \left( \frac{\sigma}{r - \Delta} \right)^{12} - \left( \frac{\sigma}{r - \Delta} \right)^{6} \right] + \varepsilon \quad & r < (r_{\mathrm{cut}} + \Delta) \\* \nonumber
                     = & 0 & r \ge (r_{\mathrm{cut}} + \Delta).
\end{eqnarray}
\normalsize

\noindent $\Delta$ is the average of particle diameters between which the potential is calculated.
$\varepsilon = 1$ for all potentials used in this study.
The pair potential parameter $h$ is used to indicate potential `hardness'.
The value of $\sigma$ used in the SWCA potential is related to $h$ as $\sigma = 1-h$.
The radial shifting of the SWCA potential is then chosen so potentials of different hardness value are zero at the same distance between particle centers (as shown in Fig.~\ref{fig:potentials}).
As $h\to 1$, $\sigma\to 0$ and the potential approaches a step function.
By parameterizing our potential space in this way, we can interpolate between atomic and colloidal-type particle interactions.
The key difference is the length scale over which the potential varies.
In atomic systems, the sharpness of potentials is fundamentally limited by natural constants and the behavior of electrons.
In colloidal systems, the pair potential values can change rapidly over distances much smaller than the particle separation distances (such as in hard-sphere systems, for instance).
We examine the extent to which these materials obey the linear elastic approximation by computing the strain fields around edge dislocations and the stress-strain relationships for homogeneous uniaxial strains.

\section{Results}

\subsection{Elastic Property Matching}

To compare repulsive and attractive potentials, a pressure (or density) must be chosen for the repulsive systems.
The Lennard-Jones (LJ) solid at zero pressure was chosen as the reference against which to match the elastic properties of the repulsive systems.
The homogeneous deformation strain energy under a collection of modes was chosen as the metric with which to compare the similarity of solids formed by different potentials.
The matching pressure is defined here as the pressure which satisfies the equation:

\begin{eqnarray}
    min \left[\sum_m \Biggl| \sum_{ijkl} \right. && C_{ijkl}^{(\textrm{LJ}, P=0)} \eta_{ij}^m \eta_{kl}^m \\* \nonumber
                                  - \sum_{ijkl} && \left. C_{ijkl}^{(\textrm{SWCA}, P=P_i)} \eta_{ij}^m \eta_{kl}^m \Biggr| \right]_{P_i}.
\end{eqnarray}

\noindent Here, $C_{ijkl}$ is the second-order elastic modulus tensor and $\eta_{ij}^m$ is a homogeneous strain.
The $\eta_{ij}^m$ (modes) over which to sum deformation energy differences could be chosen to emphasize a particular kind of deformation.
Since here we are concerned with a general measure of the mechanical similarity of two materials, we use an even weighting of all non-zero strains.
Using the six independent components of the linear strain tensor $\eta_{ij}$ and combining them with equal weight, 64 permutations are possible, with one permutation being the trivial zero strain case.

\begin{figure*}[t]
\centering
\includegraphics[width=\textwidth,keepaspectratio]{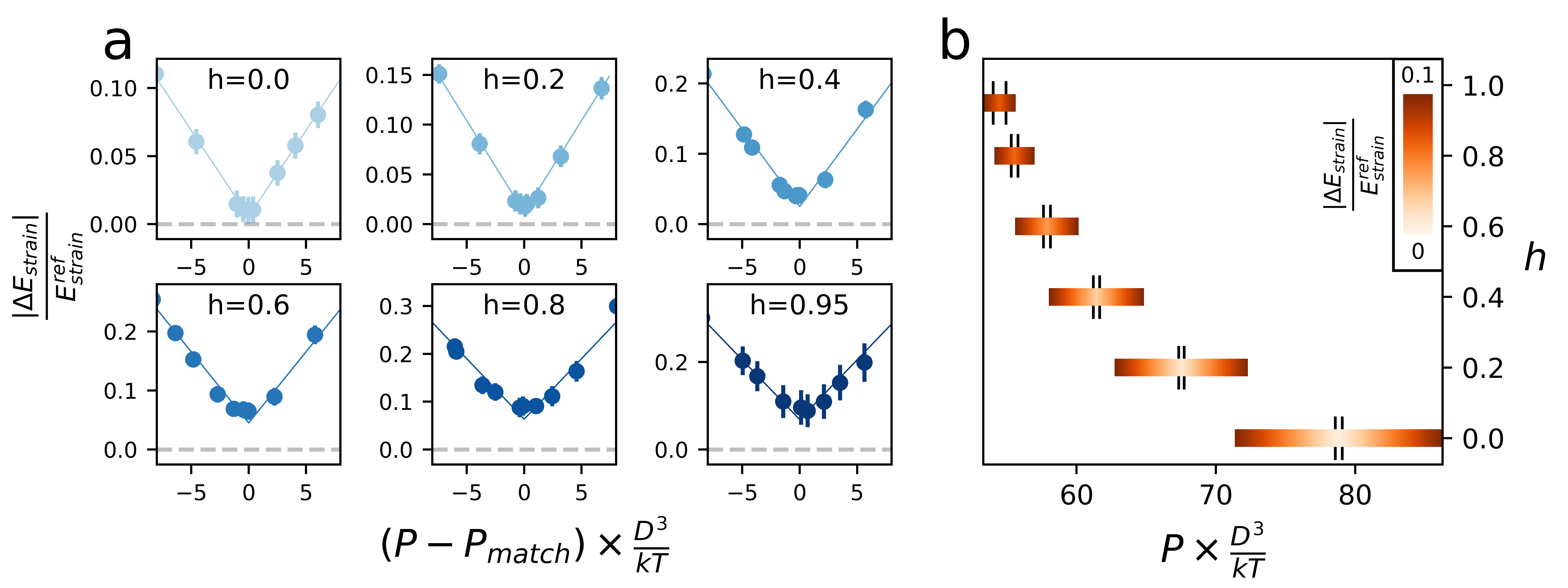}
\caption{
Pressures at which the moduli of SWCA systems most closely approach those of an LJ system at zero pressure.
\textbf{(a)} Plots of the absolute value of the fractional difference in strain energy between the SWCA and LJ system.
    Error bars are calculated \textit{via} bootstrapping of moduli measurements, and represent one standard deviation.
\textbf{(b)} Absolute pressure versus hardness.
    Tick marks indicate the uncertainty of $P_{match}$, one standard deviation of the minima from the fit in (a), and the color map is the fit value from (a), truncated at 10\% maximum difference.
}
\label{fig:pequiv}
\end{figure*}

The result of this property-matching search is shown in Fig. \ref{fig:pequiv}.
It is clear that the mechanical properties of softer potentials can be better matched to the LJ solid by varying the pressure, however it is possible to match with less than 10\% variation for all potential hardness values studied.
We observed that the pressure window of optimal matching shrinks as the potential hardness increases.

\subsection{Entropy of Deformation}

\begin{figure*}[t]
\centering
\includegraphics[width=\textwidth,keepaspectratio]{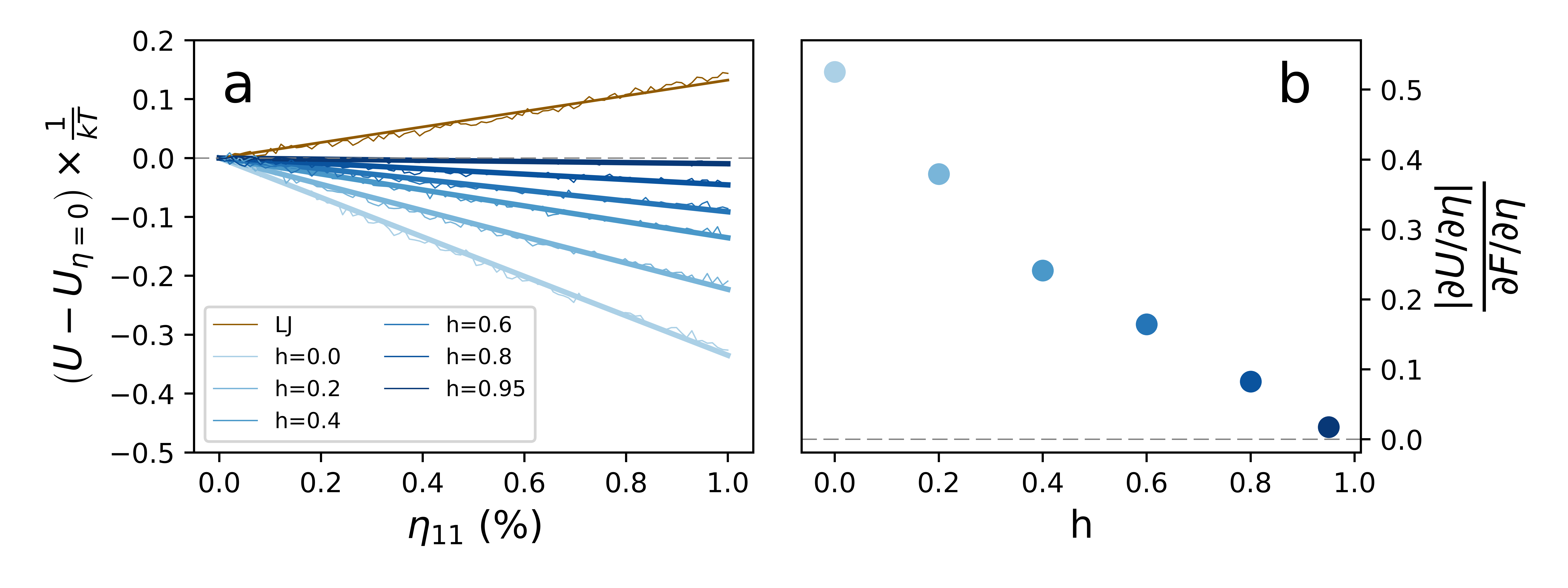}
\caption{
Energy and entropy of deformation.
\textbf{(a)} Internal (potential and kinetic) energy of deformation of potentials in this hardness family.
Solid lines are linear fits to data collected from systems homogeneously deformed along the $[1 \bar 1 0]$ direction.
\textbf{(b)} Ratio of internal energy change to free energy change (evaluated at 1\% strain).
Error bars of one standard deviation are much smaller than shown data points.
}
\label{fig:deformation_entropy}
\end{figure*}

As the hardness of the SWCA potential is increased, the behavior of the system more closely approaches that of hard spheres.
For a material of hard spheres, system configurations with particle overlaps are impossible, as these have infinite internal energy.
All other configurations have zero internal energy.
Therefore these materials are `entropic' solids, with no enthalpy changes upon crystallization.
It follows that the mechanical properties of solids studied here must also become increasingly dominated by entropy effects as $h$ approaches $1$.
Fig.~\ref{fig:deformation_entropy} demonstrates this point explicitly:
as potential hardness is increased, the internal energy of deformation decreases in magnitude, indicating that entropic changes contribute more to the free energy of deformation (and the elastic modulus) than energetic interactions between particles.
As hardness is increased, the contribution of potential energy to the free energy of deformation becomes negligible.
This demonstrates that the $h$ parameterization is useful for studying the transition from energetic deformation to entropic deformation.

\subsection{Statistics of Strain Fields in Entropic Solids}

Working with entropic materials presents a complication for analyzing strain fields: fluctuations in particle position are so large that accurate strain fields cannot be calculated from a single snapshot of a particle's local neighborhood.
In materials with energetically dominated free energies of deformation, temperature can be lowered to improve the accuracy of strain fields sampled from a single snapshot.
This is not possible for materials with significant deformation entropy.
Therefore the strain fields must be calculated from averages of particle positions.

\begin{figure}[tb]
\centering
\includegraphics[width=0.5\textwidth,keepaspectratio]{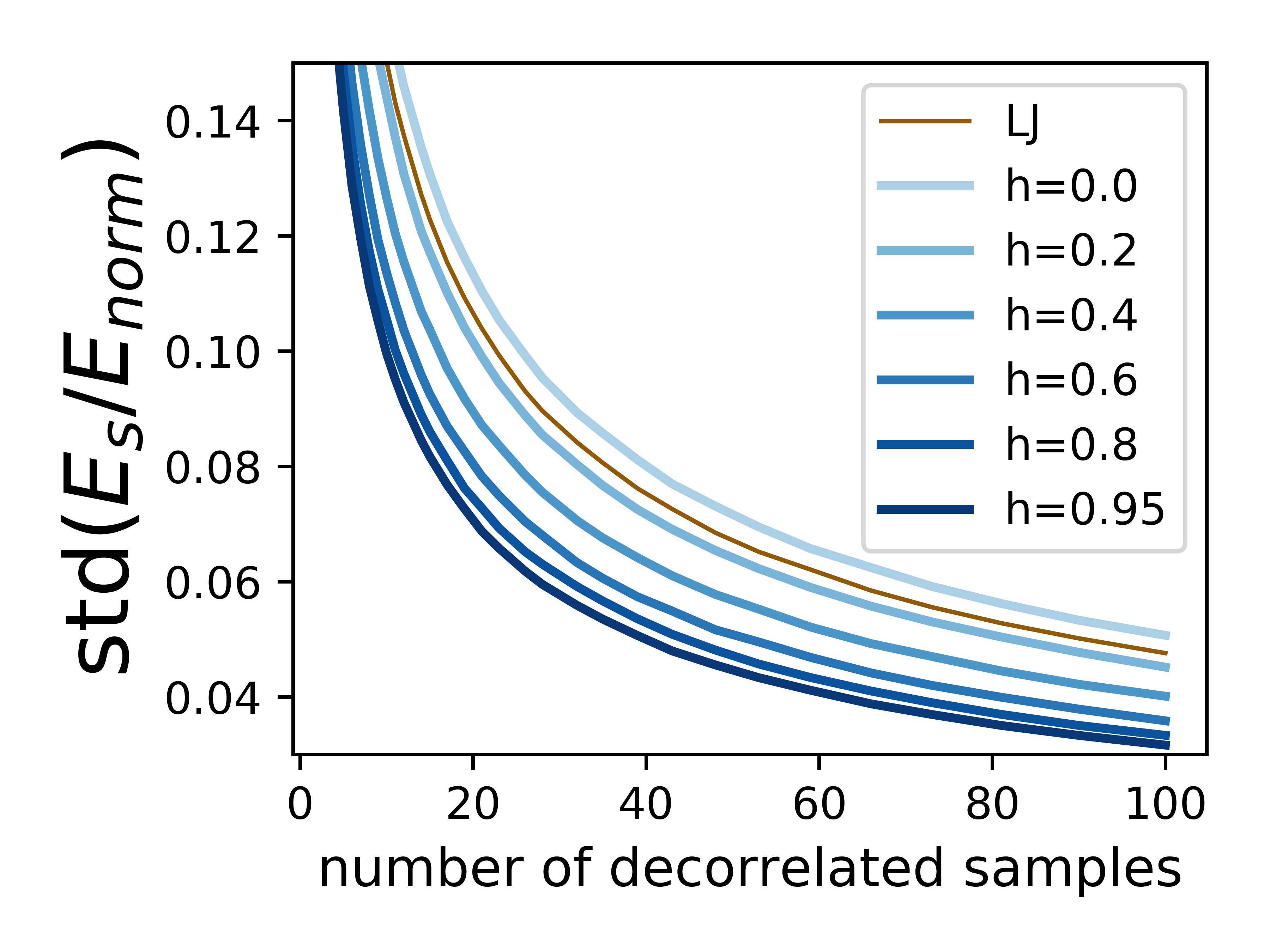}
\caption{
The standard deviation of per-particle strain energy values (normalized to a reference strain) as a function of the number of decorrelated samples considered.
For each sample number, bootstrapping is used to obtain the standard deviation.
The normalization strain is chosen to be the per-particle strain energy of a 1\% volumetric expansion in LJ, which estimates the strain energy of a dilatory point defect such as a vacancy.
With increasing $h$, fewer samples are required to obtain the same standard deviation values.
}
\label{fig:strain_convergence}
\end{figure}

To understand how the accuracy of strain field sampling changes as snapshots are collected, the per-particle strain of defect-free simulation domains were calculated with methods implemented in the visualization and analysis tool \texttt{OVITO} \cite{OVITO,Stukowski2012}.
Samples were collected with adequate time lag to assure decorrelation.
The behavior of the standard deviation of averaged measurements with the number of samples is shown in Fig.~\ref{fig:strain_convergence}.
Harder potentials converge with fewer decorrelated samples.
This is a consequence of collision behavior for different hardness potentials.
For soft potentials, velocities slow gradually as particles overlap and are eventually accelerated apart.
A hard particle ($h=1$) has instantaneous collisions.
Since collisions occur far from the average position of a particle at its lattice site, slow collisions spread the particle's probability distribution function farther from the average, \textit{i.e.}, increasing positional variance.
Hard particles with fast collisions spend less time far from their lattice sites.
Consequently, variables that are functions of particle position (such as strain) converge faster for harder particles.

\subsection{Strain Fields Around Edge Dislocations}

For all potentials in this study, dislocations initialized by the subtraction of a half-plane of particles dissociate into partial dislocations.
The easy dissociation of hard-sphere dislocations is explained by the low stacking fault energy of such systems \cite{Pronk1999}.
The per-particle strain fields, produced from 100 decorrelated samples, are shown in Fig.~\ref{fig:strain_comparison} (see Fig.~S1 in the Supplemental Material for additional strain components \cite{SupplementalMaterial}).
For comparison, analytic strain distributions calculated with the method of eigenstrains \cite{Mura} are shown for the LJ solid and the hardest potential, $h=0.95$.
Sampled dislocation strain fields closely follow those predicted by linear elasticity for LJ and $h=0$ systems.
As $h$ approaches $1$, greater asymmetry appears in the sampled distributions (see Fig.~S2 in the Supplemental Material \cite{SupplementalMaterial}).

\begin{figure*}[p]
\centering
\includegraphics[width=\textwidth,keepaspectratio]{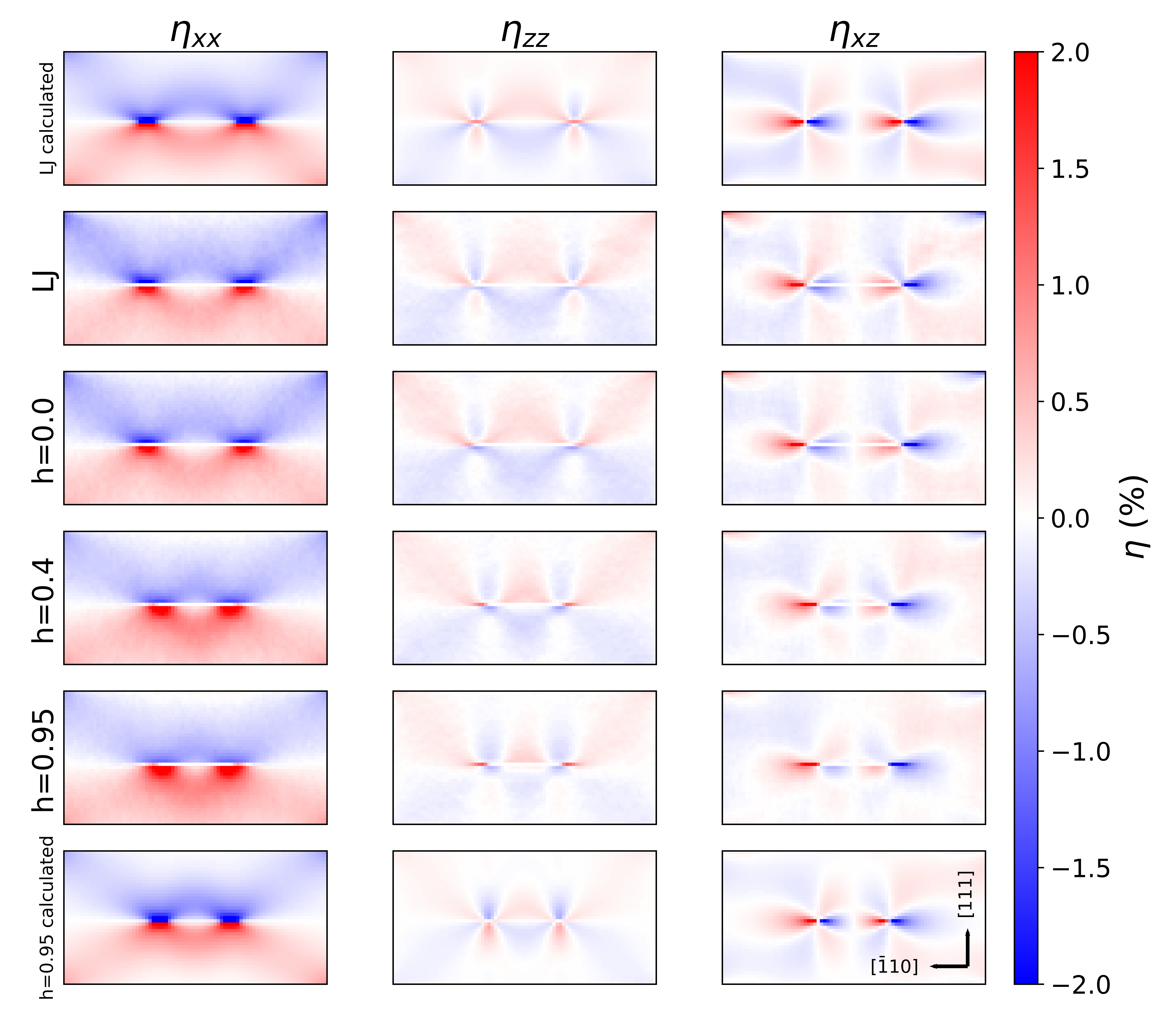}
\caption{
Strain fields around partial dislocation pairs.
This view is along the length of the dislocation line, with the Burgers vector pointing from left to right in the plane of the image.
The three strain components with the largest magnitudes are shown.
    The other three strain components contain small magnitudes because they involve the $[1 1 \bar 2]$ direction, along which there is little change in the displacement field around a straight dislocation line.
See Supplementary Material for these components \cite{SupplementalMaterial}.
At low hardness, and in particular for LJ, we find a close match between the analytically generated strain distribution and the sampled one.
As the hardness of the potential is increased, an asymmetry develops between the compressive and tensile regions of $\eta_{xx}$ and $\eta_{zz}$.
}
\label{fig:strain_comparison}
\end{figure*}

\begin{figure}[tb]
\centering
\includegraphics[width=0.5\textwidth,keepaspectratio]{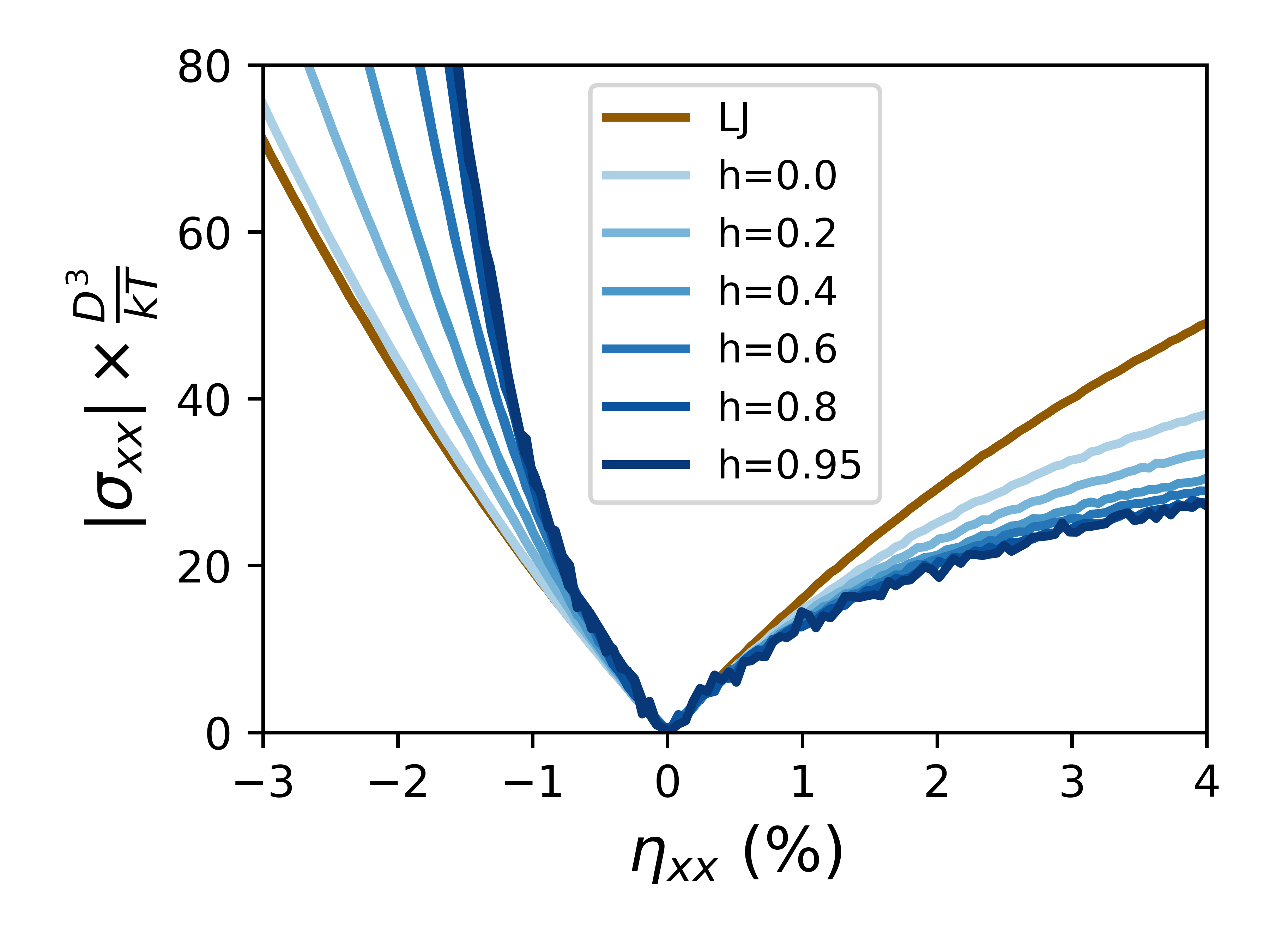}
\caption{
Stress-strain relationship for different values of $h$ under a strain ramp.
The region of linear behavior shrinks and the asymmetry of the stress curve increases as the potential hardness is increased.
}
\label{fig:strain_asymmetry}
\end{figure}

To explore the origins of strain asymmetry, we simulated defect-free systems and imposed a homogeneous uniaxial strain in the $x$ direction.
Fig.~\ref{fig:strain_asymmetry} shows the difference in stress-strain relation for the application of a compressive or tensile strain.
This result explains the asymmetry in the strain field surrounding dislocation cores.
The system can be said to be elastically linear when the stress depends linearly on strain and strain energy is symmetric for positive and negative strains.
From this plot, we can see that systems with higher $h$ have a more limited range of strains where these two conditions are true.
The LJ solid is the most symmetric, however the inherent asymmetry of the bonding well for this potential produces a difference for tension and compression at large strains (compression into the repulsive particle core increases system energy more than tension).
The pure repulsive solids all display greater stress-strain asymmetry than the LJ solid.
Furthermore, as $h$ approaches $1$, the linear strain region shrinks.

These factors together explain the strain field asymmetry of dislocations in hard-potential colloidal crystals.
The number density of dislocation lines remains the same for all systems studied here.
As the linear strain limit reduces with increasing $h$, the field close to the dislocation core begins to exceed these limits; at this point it is energetically much less costly to increase the tensile strain than the compressive strain.
As $h$ approaches $1$, the size of the region near the core that experiences nonlinear strain increases.
If the dislocation core region is defined as the region in which linear elastic theory does not accurately predict strain field magnitudes, then the core region of dislocations increases in size as the hardness of the potential is increased.
Interestingly, conventional definitions of core size, such as the distance from the core at which the shear stress is equal to the theoretical shear stress of the material \cite{Chrzan2010}, show the opposite trend (\textit{i.e.}, decreasing core size with increasing hardness).
See the Supplemental Material for details \cite{SupplementalMaterial}.

\begin{figure*}[t]
\centering
\includegraphics[width=\textwidth,keepaspectratio]{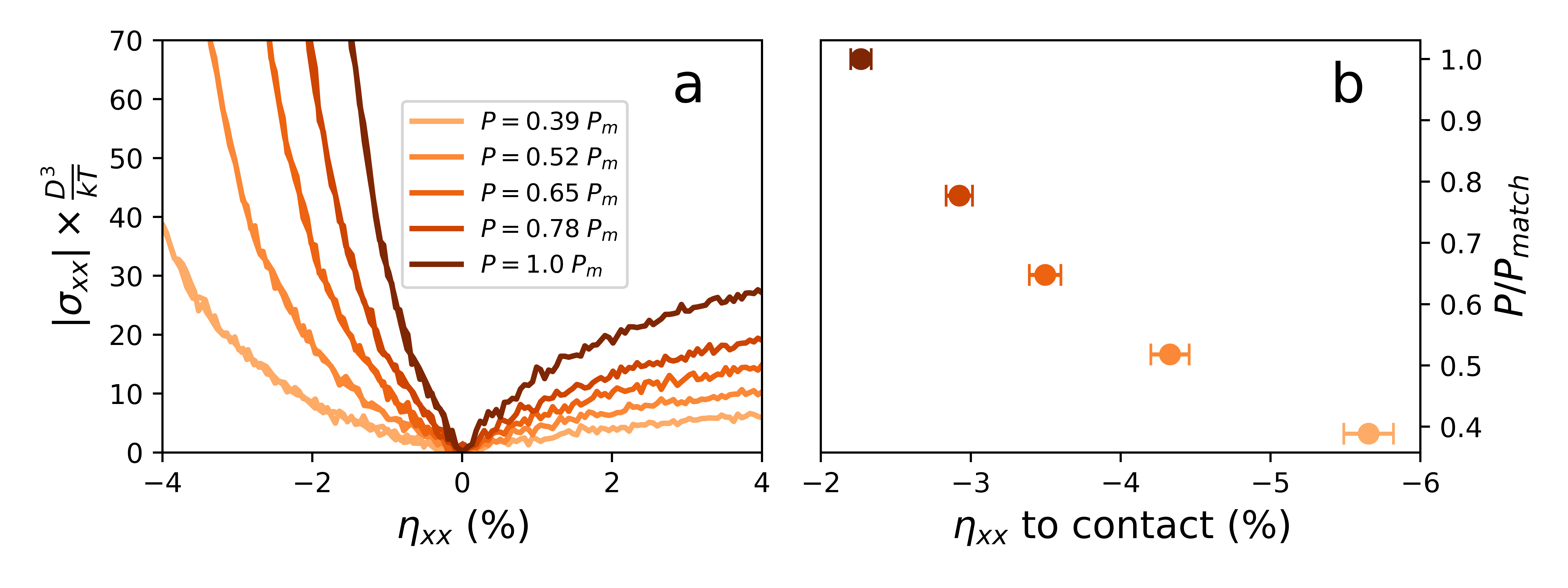}
\caption{
\textbf{(a)} Stress-strain curves for a system with $h=0.95$.
As the pressure is reduced, the linearity of the stress-strain relationship improves.
However, at pressures other than $P_{match}$, the solids cannot be said to have similar mechanical properties to the other systems shown in this study.
\textbf{(b)} Strain to contact for systems in (a).
Strain to contact is the strain needed geometrically to make particle centers sit a distance of $r_{cut}$ from each other.
    Error bars represent one standard deviation (due to uncertainty in measurement of the lattice parameter).
The region of linear behavior for a hard repulsive material must be a fraction of the strain-to-particle contact.
Together this data shows that hard potentials behave most elastically linear when the rattle volume of a particle at its lattice site is large.
The time averaged effect of many hard-particle collisions produces an effective harmonic potential.
}
\label{fig:h095_strain_asymmetry}
\end{figure*}

At pressures lower than the property matching pressure ($P_{match}$) found in Fig.~\ref{fig:pequiv}, strain asymmetry is reduced.
Fig.~\ref{fig:h095_strain_asymmetry} shows the results of the same measurement as Fig.~\ref{fig:strain_asymmetry} for particles with $h=0.95$ at a range of pressures lower than $P_{match}$.
As the pressure is lowered, the range of linear strain increases.
Fig.~\ref{fig:h095_strain_asymmetry}b shows that nonlinear strain behavior begins to occur at values that correspond to a fraction of the inter-particle spacing.
Entropic solids have inter-particle spacings significantly larger than the diameter of the particles.
Hard particles explore a volume of crystal and interact with neighbors \textit{via} collisions, which can statistically approximate harmonic lattice binding.
Hence, linear elastic behavior is recovered from hard particle solids with no explicitly defined (energetic) binding wells.
As pressure is reduced, the rattle volume of a hard particle at its lattice site increases.
Local stress results in changes to the particle rattle volume, and Fig.~\ref{fig:h095_strain_asymmetry} demonstrates that large rattle volumes tend to change more linearly with local stress changes.

\section{Conclusion}

We explored the linear elastic properties of a family of isotropic pair potentials.
The parametrization variable for the hardness of the potential, $h$, was shown to correspond to an increasing importance of entropy in the deformation free energy of colloidal crystals.
We demonstrated a method to define and search for state point conditions of maximum mechanical similarity between systems with different pair potentials.
We showed that for this family of potentials the increased importance of entropy in deformation has several consequences for strain fields:
1) faster convergence (smaller variance) of strain field fluctuations, 2) higher stress-strain asymmetry, and consequently 3) larger nonlinear strain regions around dislocation cores.
We also showed that the linear regime for these materials could be increased by lowering the pressure and thereby increasing the per-particle rattle volume.
However, this changes the mechanical properties of the solid and so precludes the matching to a reference material.

Overall, the linear elastic approximation can be usefully employed in colloidal crystals, as long as care is taken to operate within the linearity limits for the state point in question.
Applying linear elastic concepts to hard spheres at pressures significantly above the crystallization pressure will not yield good results, however at pressures near the melting point the mechanical behavior of hard spheres can be well approximated.
The LJ solid at low temperatures and zero pressure is a well-behaved harmonic solid, and the method of eigenstrains can be used to accurately predict the form of strain fields surrounding complex defects such as dislocations. The WCA potential can be matched very closely in mechanical properties to the LJ solid by the application of pressure, and the so-obtained solid is likewise well approximated by linear elastic methods.

The equivalence of complex strain fields calculated with the method of eigenstrains and the sampled strain configurations inside simulated colloidal materials opens the door to using linear elastic methods to predict and design complex strain states within colloidal materials.
Unlike in atomic materials, inclusions with designed shape can be easily introduced to colloidal crystals.
Results shown here indicate that linear elastic strain methods can be used to understand emergent interaction between defects in colloidal crystals.
So long the effects of pressure on the limits of the linear approximation is considered, the method of eigenstrains and other linear elastic methods can be powerful additions to the colloidal community's toolkit.

\section{Methods}

\subsection{Molecular Dynamics Methods}

All molecular dynamic (MD) simulations reported here are performed with \texttt{HOOMD-blue} \cite{Anderson2008, Glaser2015}.
Particles are simulated in a periodic box with Cartesian directions $x$, $y$, $z$ aligned to crystal directions $[1\bar 1 0]$, $[1 1 \bar 2]$, $[111]$ in the face-centered cubic (FCC) structure.
Simulations are carried out in the $NPT$ ensemble \textit{via} equations derived by Martyna \textit{et al.} \cite{Martyna1996}.

\subsection{Strain Fields of Dislocations in Continuum Materials}

The analytical form of the strain distribution surrounding a dislocation can be most easily found with the method of eigenstrains derived by Mura \cite{Mura}.
In that approach, the distortions from the defect are introduced as singularities and regularized by the application of a function constructed from the elastic modulus tensor of the material in the Fourier domain.
The result is a continuous periodic strain distribution.

Strain distributions for edge dislocations are calculated from the eigenstrain procedure by introducing a region of $\eta_{xx}$ strain bounded by a bi-variate Gaussian in the $x$ and $z$ directions and constant in the $y$ direction (along the dislocation line).
The center of the eigenstrain region contains a discontinuity, which defines the location of the dislocation core.
This procedure is an adaptation of methods used by Chiu \cite{Chiu1977} and Daw \cite{Daw2006}.

\subsection{Elastic Moduli Sampling}

The eigenstrain calculation has only a single material-dependent input, the elastic modulus tensor.
We sample the elastic modulus tensor from MD simulations following the method of Gusev \textit{et al.}~\cite{Gusev1996}: moduli are calculated from the fluctuations of virial stress and box strain for an MD simulation which permits the box geometry to change ($NPT$ ensemble).
A constant cutoff $r_{\mathrm{cut}} = 3 \sigma \cdot 2^{1/6}$ is used for the LJ system.
At this range, the potential value is $\leq 0.3 \%$ of the attractive well depth.
The potential is shifted and a smoothing function applied so that potential energy and force are zero at $r=r_{\mathrm{cut}}$.
A $\sigma$ value of 1 is used for the LJ potential.
SWCA shifting parameters ($\sigma$ and $\Delta$) are chosen so that all potentials are zero at $r=2^{\frac{1}{6}}$.

\subsection{Strain Field Sampling}

The strain fields surrounding dislocations were sampled by histogramming per-particle strain matrices.
Per-particle strain was calculated using the analysis software \texttt{OVITO} \cite{OVITO}.
To map the per-particle data onto the output of the eigenstrain calculation, the individual strain tensors were assigned to spatial bins and averaged over time.
The number of bins was chosen to be similar to the number of particles, as this yielded the highest-resolution voxelized strain field data.
The mean and standard deviation of the mean for per-particle strain fields were calculated with  bootstrapping \cite{Efron1994}.

\begin{acknowledgments}
    This research was supported by the National Science Foundation, Division of Materials Research Award \# DMR 1409620.
    Computational resources and services were supported by Advanced Research Computing at the University of Michigan, Ann Arbor.
    We thank Dr.\ L.\ Qi for helpful discussions.
\end{acknowledgments}


\begin{thebibliography}{43}%
\makeatletter
\providecommand \@ifxundefined [1]{%
 \@ifx{#1\undefined}
}%
\providecommand \@ifnum [1]{%
 \ifnum #1\expandafter \@firstoftwo
 \else \expandafter \@secondoftwo
 \fi
}%
\providecommand \@ifx [1]{%
 \ifx #1\expandafter \@firstoftwo
 \else \expandafter \@secondoftwo
 \fi
}%
\providecommand \natexlab [1]{#1}%
\providecommand \enquote  [1]{``#1''}%
\providecommand \bibnamefont  [1]{#1}%
\providecommand \bibfnamefont [1]{#1}%
\providecommand \citenamefont [1]{#1}%
\providecommand \href@noop [0]{\@secondoftwo}%
\providecommand \href [0]{\begingroup \@sanitize@url \@href}%
\providecommand \@href[1]{\@@startlink{#1}\@@href}%
\providecommand \@@href[1]{\endgroup#1\@@endlink}%
\providecommand \@sanitize@url [0]{\catcode `\\12\catcode `\$12\catcode
  `\&12\catcode `\#12\catcode `\^12\catcode `\_12\catcode `\%12\relax}%
\providecommand \@@startlink[1]{}%
\providecommand \@@endlink[0]{}%
\providecommand \url  [0]{\begingroup\@sanitize@url \@url }%
\providecommand \@url [1]{\endgroup\@href {#1}{\urlprefix }}%
\providecommand \urlprefix  [0]{URL }%
\providecommand \Eprint [0]{\href }%
\providecommand \doibase [0]{http://dx.doi.org/}%
\providecommand \selectlanguage [0]{\@gobble}%
\providecommand \bibinfo  [0]{\@secondoftwo}%
\providecommand \bibfield  [0]{\@secondoftwo}%
\providecommand \translation [1]{[#1]}%
\providecommand \BibitemOpen [0]{}%
\providecommand \bibitemStop [0]{}%
\providecommand \bibitemNoStop [0]{.\EOS\space}%
\providecommand \EOS [0]{\spacefactor3000\relax}%
\providecommand \BibitemShut  [1]{\csname bibitem#1\endcsname}%
\let\auto@bib@innerbib\@empty
\bibitem [{\citenamefont {Glotzer}\ and\ \citenamefont
  {Solomon}(2007)}]{Glotzer2007}%
  \BibitemOpen
  \bibfield  {author} {\bibinfo {author} {\bibfnamefont {S.~C.}\ \bibnamefont
  {Glotzer}}\ and\ \bibinfo {author} {\bibfnamefont {M.~J.}\ \bibnamefont
  {Solomon}},\ }\href {\doibase 10.1038/nmat1949} {\bibfield  {journal}
  {\bibinfo  {journal} {Nature Materials}\ }\textbf {\bibinfo {volume} {6}},\
  \bibinfo {pages} {557} (\bibinfo {year} {2007})}\BibitemShut {NoStop}%
\bibitem [{\citenamefont {Nykypanchuk}\ \emph {et~al.}(2008)\citenamefont
  {Nykypanchuk}, \citenamefont {Maye}, \citenamefont {Lelie},\ and\
  \citenamefont {Gang}}]{Nykypanchuk2008}%
  \BibitemOpen
  \bibfield  {author} {\bibinfo {author} {\bibfnamefont {D.}~\bibnamefont
  {Nykypanchuk}}, \bibinfo {author} {\bibfnamefont {M.~M.}\ \bibnamefont
  {Maye}}, \bibinfo {author} {\bibfnamefont {D.~v.~d.}\ \bibnamefont {Lelie}},
  \ and\ \bibinfo {author} {\bibfnamefont {O.}~\bibnamefont {Gang}},\ }\href
  {\doibase 10.1038/nature06560} {\bibfield  {journal} {\bibinfo  {journal}
  {Nature}\ }\textbf {\bibinfo {volume} {451}},\ \bibinfo {pages} {549}
  (\bibinfo {year} {2008})}\BibitemShut {NoStop}%
\bibitem [{\citenamefont {Sacanna}\ and\ \citenamefont
  {Pine}(2011)}]{Sacanna2011}%
  \BibitemOpen
  \bibfield  {author} {\bibinfo {author} {\bibfnamefont {S.}~\bibnamefont
  {Sacanna}}\ and\ \bibinfo {author} {\bibfnamefont {D.~J.}\ \bibnamefont
  {Pine}},\ }\href {\doibase 10.1016/j.cocis.2011.01.003} {\bibfield  {journal}
  {\bibinfo  {journal} {Current Opinion in Colloid \& Interface Science}\
  }\textbf {\bibinfo {volume} {16}},\ \bibinfo {pages} {96} (\bibinfo {year}
  {2011})}\BibitemShut {NoStop}%
\bibitem [{\citenamefont {Haji-Akbari}\ \emph {et~al.}(2009)\citenamefont
  {Haji-Akbari}, \citenamefont {Engel}, \citenamefont {Keys}, \citenamefont
  {Zheng}, \citenamefont {Petschek}, \citenamefont {Palffy-Muhoray},\ and\
  \citenamefont {Glotzer}}]{HajiAkbari2009}%
  \BibitemOpen
  \bibfield  {author} {\bibinfo {author} {\bibfnamefont {A.}~\bibnamefont
  {Haji-Akbari}}, \bibinfo {author} {\bibfnamefont {M.}~\bibnamefont {Engel}},
  \bibinfo {author} {\bibfnamefont {A.~S.}\ \bibnamefont {Keys}}, \bibinfo
  {author} {\bibfnamefont {X.}~\bibnamefont {Zheng}}, \bibinfo {author}
  {\bibfnamefont {R.~G.}\ \bibnamefont {Petschek}}, \bibinfo {author}
  {\bibfnamefont {P.}~\bibnamefont {Palffy-Muhoray}}, \ and\ \bibinfo {author}
  {\bibfnamefont {S.~C.}\ \bibnamefont {Glotzer}},\ }\href {\doibase
  10.1038/nature08641} {\bibfield  {journal} {\bibinfo  {journal} {Nature}\
  }\textbf {\bibinfo {volume} {462}},\ \bibinfo {pages} {773} (\bibinfo {year}
  {2009})}\BibitemShut {NoStop}%
\bibitem [{\citenamefont {Wang}\ \emph {et~al.}(2015)\citenamefont {Wang},
  \citenamefont {Wang}, \citenamefont {Zheng}, \citenamefont {Ducrot},
  \citenamefont {Yodh}, \citenamefont {Weck},\ and\ \citenamefont
  {Pine}}]{Wang2015}%
  \BibitemOpen
  \bibfield  {author} {\bibinfo {author} {\bibfnamefont {Y.}~\bibnamefont
  {Wang}}, \bibinfo {author} {\bibfnamefont {Y.}~\bibnamefont {Wang}}, \bibinfo
  {author} {\bibfnamefont {X.}~\bibnamefont {Zheng}}, \bibinfo {author}
  {\bibfnamefont {{\'E}.}~\bibnamefont {Ducrot}}, \bibinfo {author}
  {\bibfnamefont {J.~S.}\ \bibnamefont {Yodh}}, \bibinfo {author}
  {\bibfnamefont {M.}~\bibnamefont {Weck}}, \ and\ \bibinfo {author}
  {\bibfnamefont {D.~J.}\ \bibnamefont {Pine}},\ }\href {\doibase
  10.1038/ncomms8253} {\bibfield  {journal} {\bibinfo  {journal} {Nature
  Communications}\ }\textbf {\bibinfo {volume} {6}},\ \bibinfo {pages} {7253}
  (\bibinfo {year} {2015})}\BibitemShut {NoStop}%
\bibitem [{\citenamefont {Ducrot}\ \emph {et~al.}(2017)\citenamefont {Ducrot},
  \citenamefont {He}, \citenamefont {Yi},\ and\ \citenamefont
  {Pine}}]{Ducrot2017}%
  \BibitemOpen
  \bibfield  {author} {\bibinfo {author} {\bibfnamefont {{\'E}.}~\bibnamefont
  {Ducrot}}, \bibinfo {author} {\bibfnamefont {M.}~\bibnamefont {He}}, \bibinfo
  {author} {\bibfnamefont {G.-R.}\ \bibnamefont {Yi}}, \ and\ \bibinfo {author}
  {\bibfnamefont {D.~J.}\ \bibnamefont {Pine}},\ }\href {\doibase
  10.1038/nmat4869} {\bibfield  {journal} {\bibinfo  {journal} {Nature
  Materials}\ }\textbf {\bibinfo {volume} {16}},\ \bibinfo {pages} {652}
  (\bibinfo {year} {2017})}\BibitemShut {NoStop}%
\bibitem [{\citenamefont {Chen}\ \emph {et~al.}(2011)\citenamefont {Chen},
  \citenamefont {Bae},\ and\ \citenamefont {Granick}}]{Chen2011}%
  \BibitemOpen
  \bibfield  {author} {\bibinfo {author} {\bibfnamefont {Q.}~\bibnamefont
  {Chen}}, \bibinfo {author} {\bibfnamefont {S.~C.}\ \bibnamefont {Bae}}, \
  and\ \bibinfo {author} {\bibfnamefont {S.}~\bibnamefont {Granick}},\ }\href
  {\doibase 10.1038/nature09713} {\bibfield  {journal} {\bibinfo  {journal}
  {Nature}\ }\textbf {\bibinfo {volume} {469}},\ \bibinfo {pages} {381}
  (\bibinfo {year} {2011})}\BibitemShut {NoStop}%
\bibitem [{\citenamefont {Damasceno}\ \emph {et~al.}(2012)\citenamefont
  {Damasceno}, \citenamefont {Engel},\ and\ \citenamefont
  {Glotzer}}]{Damasceno2012}%
  \BibitemOpen
  \bibfield  {author} {\bibinfo {author} {\bibfnamefont {P.~F.}\ \bibnamefont
  {Damasceno}}, \bibinfo {author} {\bibfnamefont {M.}~\bibnamefont {Engel}}, \
  and\ \bibinfo {author} {\bibfnamefont {S.~C.}\ \bibnamefont {Glotzer}},\
  }\href {\doibase 10.1126/science.1220869} {\bibfield  {journal} {\bibinfo
  {journal} {Science}\ }\textbf {\bibinfo {volume} {337}},\ \bibinfo {pages}
  {453} (\bibinfo {year} {2012})}\BibitemShut {NoStop}%
\bibitem [{\citenamefont {Zhang}\ \emph {et~al.}(2013)\citenamefont {Zhang},
  \citenamefont {Lu}, \citenamefont {Yager}, \citenamefont {Lelie},\ and\
  \citenamefont {Gang}}]{Zhang2013}%
  \BibitemOpen
  \bibfield  {author} {\bibinfo {author} {\bibfnamefont {Y.}~\bibnamefont
  {Zhang}}, \bibinfo {author} {\bibfnamefont {F.}~\bibnamefont {Lu}}, \bibinfo
  {author} {\bibfnamefont {K.~G.}\ \bibnamefont {Yager}}, \bibinfo {author}
  {\bibfnamefont {D.~v.~d.}\ \bibnamefont {Lelie}}, \ and\ \bibinfo {author}
  {\bibfnamefont {O.}~\bibnamefont {Gang}},\ }\href {\doibase
  10.1038/nnano.2013.209} {\bibfield  {journal} {\bibinfo  {journal} {Nature
  Nanotechnology}\ }\textbf {\bibinfo {volume} {8}},\ \bibinfo {pages} {865}
  (\bibinfo {year} {2013})}\BibitemShut {NoStop}%
\bibitem [{\citenamefont {van Anders}\ \emph {et~al.}(2015)\citenamefont {van
  Anders}, \citenamefont {Klotsa}, \citenamefont {Karas}, \citenamefont
  {Dodd},\ and\ \citenamefont {Glotzer}}]{vanAnders2015}%
  \BibitemOpen
  \bibfield  {author} {\bibinfo {author} {\bibfnamefont {G.}~\bibnamefont {van
  Anders}}, \bibinfo {author} {\bibfnamefont {D.}~\bibnamefont {Klotsa}},
  \bibinfo {author} {\bibfnamefont {A.~S.}\ \bibnamefont {Karas}}, \bibinfo
  {author} {\bibfnamefont {P.~M.}\ \bibnamefont {Dodd}}, \ and\ \bibinfo
  {author} {\bibfnamefont {S.~C.}\ \bibnamefont {Glotzer}},\ }\href {\doibase
  10.1021/acsnano.5b04181} {\bibfield  {journal} {\bibinfo  {journal} {ACS
  Nano}\ }\textbf {\bibinfo {volume} {9}},\ \bibinfo {pages} {9542} (\bibinfo
  {year} {2015})}\BibitemShut {NoStop}%
\bibitem [{\citenamefont {Kumar}\ \emph {et~al.}(2017)\citenamefont {Kumar},
  \citenamefont {Kumaraswamy}, \citenamefont {Prasad}, \citenamefont
  {Bandyopadhyaya}, \citenamefont {Granick}, \citenamefont {Gang},
  \citenamefont {Manoharan}, \citenamefont {Frenkel},\ and\ \citenamefont
  {Kotov}}]{Kumar2017}%
  \BibitemOpen
  \bibfield  {author} {\bibinfo {author} {\bibfnamefont {S.~K.}\ \bibnamefont
  {Kumar}}, \bibinfo {author} {\bibfnamefont {G.}~\bibnamefont {Kumaraswamy}},
  \bibinfo {author} {\bibfnamefont {B.~L.~V.}\ \bibnamefont {Prasad}}, \bibinfo
  {author} {\bibfnamefont {R.}~\bibnamefont {Bandyopadhyaya}}, \bibinfo
  {author} {\bibfnamefont {S.}~\bibnamefont {Granick}}, \bibinfo {author}
  {\bibfnamefont {O.}~\bibnamefont {Gang}}, \bibinfo {author} {\bibfnamefont
  {V.~N.}\ \bibnamefont {Manoharan}}, \bibinfo {author} {\bibfnamefont
  {D.}~\bibnamefont {Frenkel}}, \ and\ \bibinfo {author} {\bibfnamefont
  {N.~A.}\ \bibnamefont {Kotov}},\ }\href {\doibase
  10.18520/cs/v112/i08/1635-1641} {\bibfield  {journal} {\bibinfo  {journal}
  {Current Science}\ }\textbf {\bibinfo {volume} {112}},\ \bibinfo {pages}
  {1635} (\bibinfo {year} {2017})}\BibitemShut {NoStop}%
\bibitem [{\citenamefont {Schall}\ \emph {et~al.}(2004)\citenamefont {Schall},
  \citenamefont {Cohen}, \citenamefont {Weitz},\ and\ \citenamefont
  {Spaepen}}]{Schall2004}%
  \BibitemOpen
  \bibfield  {author} {\bibinfo {author} {\bibfnamefont {P.}~\bibnamefont
  {Schall}}, \bibinfo {author} {\bibfnamefont {I.}~\bibnamefont {Cohen}},
  \bibinfo {author} {\bibfnamefont {D.~A.}\ \bibnamefont {Weitz}}, \ and\
  \bibinfo {author} {\bibfnamefont {F.}~\bibnamefont {Spaepen}},\ }\href
  {\doibase 10.1126/science.1102186} {\bibfield  {journal} {\bibinfo  {journal}
  {Science}\ }\textbf {\bibinfo {volume} {305}},\ \bibinfo {pages} {1944}
  (\bibinfo {year} {2004})}\BibitemShut {NoStop}%
\bibitem [{\citenamefont {Schall}\ \emph {et~al.}(2006)\citenamefont {Schall},
  \citenamefont {Cohen}, \citenamefont {Weitz},\ and\ \citenamefont
  {Spaepen}}]{Schall2006}%
  \BibitemOpen
  \bibfield  {author} {\bibinfo {author} {\bibfnamefont {P.}~\bibnamefont
  {Schall}}, \bibinfo {author} {\bibfnamefont {I.}~\bibnamefont {Cohen}},
  \bibinfo {author} {\bibfnamefont {D.~A.}\ \bibnamefont {Weitz}}, \ and\
  \bibinfo {author} {\bibfnamefont {F.}~\bibnamefont {Spaepen}},\ }\href
  {\doibase 10.1038/nature04557} {\bibfield  {journal} {\bibinfo  {journal}
  {Nature}\ }\textbf {\bibinfo {volume} {440}},\ \bibinfo {pages} {319}
  (\bibinfo {year} {2006})}\BibitemShut {NoStop}%
\bibitem [{\citenamefont {Lin}\ \emph {et~al.}(2016)\citenamefont {Lin},
  \citenamefont {Bierbaum}, \citenamefont {Schall}, \citenamefont {Sethna},\
  and\ \citenamefont {Cohen}}]{Lin2016}%
  \BibitemOpen
  \bibfield  {author} {\bibinfo {author} {\bibfnamefont {N.~Y.~C.}\
  \bibnamefont {Lin}}, \bibinfo {author} {\bibfnamefont {M.}~\bibnamefont
  {Bierbaum}}, \bibinfo {author} {\bibfnamefont {P.}~\bibnamefont {Schall}},
  \bibinfo {author} {\bibfnamefont {J.~P.}\ \bibnamefont {Sethna}}, \ and\
  \bibinfo {author} {\bibfnamefont {I.}~\bibnamefont {Cohen}},\ }\href
  {\doibase 10.1038/nmat4715} {\bibfield  {journal} {\bibinfo  {journal}
  {Nature Materials}\ }\textbf {\bibinfo {volume} {15}},\ \bibinfo {pages}
  {1172} (\bibinfo {year} {2016})}\BibitemShut {NoStop}%
\bibitem [{\citenamefont {van~der Meer}\ \emph {et~al.}(2017)\citenamefont
  {van~der Meer}, \citenamefont {Dijkstra},\ and\ \citenamefont
  {Filion}}]{vdMeer2017}%
  \BibitemOpen
  \bibfield  {author} {\bibinfo {author} {\bibfnamefont {B.}~\bibnamefont
  {van~der Meer}}, \bibinfo {author} {\bibfnamefont {M.}~\bibnamefont
  {Dijkstra}}, \ and\ \bibinfo {author} {\bibfnamefont {L.}~\bibnamefont
  {Filion}},\ }\href {\doibase 10.1063/1.4990416} {\bibfield  {journal}
  {\bibinfo  {journal} {The Journal of Chemical Physics}\ }\textbf {\bibinfo
  {volume} {146}},\ \bibinfo {pages} {244905} (\bibinfo {year}
  {2017})}\BibitemShut {NoStop}%
\bibitem [{\citenamefont {Yablonovitch}(1987)}]{Yablonovitch1987}%
  \BibitemOpen
  \bibfield  {author} {\bibinfo {author} {\bibfnamefont {E.}~\bibnamefont
  {Yablonovitch}},\ }\href {\doibase 10.1103/PhysRevLett.58.2059} {\bibfield
  {journal} {\bibinfo  {journal} {Physical Review Letters}\ }\textbf {\bibinfo
  {volume} {58}},\ \bibinfo {pages} {2059} (\bibinfo {year}
  {1987})}\BibitemShut {NoStop}%
\bibitem [{\citenamefont {Yablonovitch}\ \emph {et~al.}(1991)\citenamefont
  {Yablonovitch}, \citenamefont {Gmitter},\ and\ \citenamefont
  {Leung}}]{Yablonovitch1991}%
  \BibitemOpen
  \bibfield  {author} {\bibinfo {author} {\bibfnamefont {E.}~\bibnamefont
  {Yablonovitch}}, \bibinfo {author} {\bibfnamefont {T.~J.}\ \bibnamefont
  {Gmitter}}, \ and\ \bibinfo {author} {\bibfnamefont {K.~M.}\ \bibnamefont
  {Leung}},\ }\href {\doibase 10.1103/PhysRevLett.67.2295} {\bibfield
  {journal} {\bibinfo  {journal} {Physical Review Letters}\ }\textbf {\bibinfo
  {volume} {67}},\ \bibinfo {pages} {2295} (\bibinfo {year}
  {1991})}\BibitemShut {NoStop}%
\bibitem [{\citenamefont {Maldovan}\ and\ \citenamefont
  {Thomas}(2004)}]{Maldovan2004}%
  \BibitemOpen
  \bibfield  {author} {\bibinfo {author} {\bibfnamefont {M.}~\bibnamefont
  {Maldovan}}\ and\ \bibinfo {author} {\bibfnamefont {E.~L.}\ \bibnamefont
  {Thomas}},\ }\href {\doibase 10.1038/nmat1201} {\bibfield  {journal}
  {\bibinfo  {journal} {Nature Materials}\ }\textbf {\bibinfo {volume} {3}},\
  \bibinfo {pages} {593} (\bibinfo {year} {2004})}\BibitemShut {NoStop}%
\bibitem [{\citenamefont {Ho}\ \emph {et~al.}(1990)\citenamefont {Ho},
  \citenamefont {Chan},\ and\ \citenamefont {Soukoulis}}]{Ho1990}%
  \BibitemOpen
  \bibfield  {author} {\bibinfo {author} {\bibfnamefont {K.~M.}\ \bibnamefont
  {Ho}}, \bibinfo {author} {\bibfnamefont {C.~T.}\ \bibnamefont {Chan}}, \ and\
  \bibinfo {author} {\bibfnamefont {C.~M.}\ \bibnamefont {Soukoulis}},\ }\href
  {\doibase 10.1103/PhysRevLett.65.3152} {\bibfield  {journal} {\bibinfo
  {journal} {Physical Review Letters}\ }\textbf {\bibinfo {volume} {65}},\
  \bibinfo {pages} {3152} (\bibinfo {year} {1990})}\BibitemShut {NoStop}%
\bibitem [{\citenamefont {Kim}\ \emph {et~al.}(2016)\citenamefont {Kim},
  \citenamefont {Macfarlane}, \citenamefont {Jones},\ and\ \citenamefont
  {Mirkin}}]{Kim2016}%
  \BibitemOpen
  \bibfield  {author} {\bibinfo {author} {\bibfnamefont {Y.}~\bibnamefont
  {Kim}}, \bibinfo {author} {\bibfnamefont {R.~J.}\ \bibnamefont {Macfarlane}},
  \bibinfo {author} {\bibfnamefont {M.~R.}\ \bibnamefont {Jones}}, \ and\
  \bibinfo {author} {\bibfnamefont {C.~A.}\ \bibnamefont {Mirkin}},\ }\href
  {\doibase 10.1126/science.aad2212} {\bibfield  {journal} {\bibinfo  {journal}
  {Science}\ }\textbf {\bibinfo {volume} {351}},\ \bibinfo {pages} {579}
  (\bibinfo {year} {2016})}\BibitemShut {NoStop}%
\bibitem [{\citenamefont {Tretiakov}\ and\ \citenamefont
  {Wojciechowski}(2014)}]{Tretiakov2014}%
  \BibitemOpen
  \bibfield  {author} {\bibinfo {author} {\bibfnamefont {K.~V.}\ \bibnamefont
  {Tretiakov}}\ and\ \bibinfo {author} {\bibfnamefont {K.~W.}\ \bibnamefont
  {Wojciechowski}},\ }\href {\doibase 10.1002/pssb.201384244} {\bibfield
  {journal} {\bibinfo  {journal} {physica status solidi (b)}\ }\textbf
  {\bibinfo {volume} {251}},\ \bibinfo {pages} {383} (\bibinfo {year}
  {2014})}\BibitemShut {NoStop}%
\bibitem [{\citenamefont {Zaccarelli}\ \emph {et~al.}(2001)\citenamefont
  {Zaccarelli}, \citenamefont {Foffi}, \citenamefont {Dawson}, \citenamefont
  {Sciortino},\ and\ \citenamefont {Tartaglia}}]{Zaccarelli2001}%
  \BibitemOpen
  \bibfield  {author} {\bibinfo {author} {\bibfnamefont {E.}~\bibnamefont
  {Zaccarelli}}, \bibinfo {author} {\bibfnamefont {G.}~\bibnamefont {Foffi}},
  \bibinfo {author} {\bibfnamefont {K.~A.}\ \bibnamefont {Dawson}}, \bibinfo
  {author} {\bibfnamefont {F.}~\bibnamefont {Sciortino}}, \ and\ \bibinfo
  {author} {\bibfnamefont {P.}~\bibnamefont {Tartaglia}},\ }\href {\doibase
  10.1103/PhysRevE.63.031501} {\bibfield  {journal} {\bibinfo  {journal}
  {Physical Review E}\ }\textbf {\bibinfo {volume} {63}},\ \bibinfo {pages}
  {031501} (\bibinfo {year} {2001})}\BibitemShut {NoStop}%
\bibitem [{\citenamefont {Zaccone}\ \emph {et~al.}(2009)\citenamefont
  {Zaccone}, \citenamefont {Wu},\ and\ \citenamefont {Del~Gado}}]{Zaccone2009}%
  \BibitemOpen
  \bibfield  {author} {\bibinfo {author} {\bibfnamefont {A.}~\bibnamefont
  {Zaccone}}, \bibinfo {author} {\bibfnamefont {H.}~\bibnamefont {Wu}}, \ and\
  \bibinfo {author} {\bibfnamefont {E.}~\bibnamefont {Del~Gado}},\ }\href
  {\doibase 10.1103/PhysRevLett.103.208301} {\bibfield  {journal} {\bibinfo
  {journal} {Physical Review Letters}\ }\textbf {\bibinfo {volume} {103}},\
  \bibinfo {pages} {208301} (\bibinfo {year} {2009})}\BibitemShut {NoStop}%
\bibitem [{\citenamefont {Grason}(2016)}]{Grason2016}%
  \BibitemOpen
  \bibfield  {author} {\bibinfo {author} {\bibfnamefont {G.~M.}\ \bibnamefont
  {Grason}},\ }\href {\doibase 10.1063/1.4962629} {\bibfield  {journal}
  {\bibinfo  {journal} {The Journal of Chemical Physics}\ }\textbf {\bibinfo
  {volume} {145}},\ \bibinfo {pages} {110901} (\bibinfo {year}
  {2016})}\BibitemShut {NoStop}%
\bibitem [{\citenamefont {Irvine}\ \emph {et~al.}(2010)\citenamefont {Irvine},
  \citenamefont {Vitelli},\ and\ \citenamefont {Chaikin}}]{Irvine2010}%
  \BibitemOpen
  \bibfield  {author} {\bibinfo {author} {\bibfnamefont {W.~T.~M.}\
  \bibnamefont {Irvine}}, \bibinfo {author} {\bibfnamefont {V.}~\bibnamefont
  {Vitelli}}, \ and\ \bibinfo {author} {\bibfnamefont {P.~M.}\ \bibnamefont
  {Chaikin}},\ }\href {\doibase 10.1038/nature09620} {\bibfield  {journal}
  {\bibinfo  {journal} {Nature}\ }\textbf {\bibinfo {volume} {468}},\ \bibinfo
  {pages} {947} (\bibinfo {year} {2010})}\BibitemShut {NoStop}%
\bibitem [{\citenamefont {Meng}\ \emph {et~al.}(2014)\citenamefont {Meng},
  \citenamefont {Paulose}, \citenamefont {Nelson},\ and\ \citenamefont
  {Manoharan}}]{Meng2014}%
  \BibitemOpen
  \bibfield  {author} {\bibinfo {author} {\bibfnamefont {G.}~\bibnamefont
  {Meng}}, \bibinfo {author} {\bibfnamefont {J.}~\bibnamefont {Paulose}},
  \bibinfo {author} {\bibfnamefont {D.~R.}\ \bibnamefont {Nelson}}, \ and\
  \bibinfo {author} {\bibfnamefont {V.~N.}\ \bibnamefont {Manoharan}},\ }\href
  {\doibase 10.1126/science.1244827} {\bibfield  {journal} {\bibinfo  {journal}
  {Science}\ }\textbf {\bibinfo {volume} {343}},\ \bibinfo {pages} {634}
  (\bibinfo {year} {2014})}\BibitemShut {NoStop}%
\bibitem [{\citenamefont {Azadi}\ and\ \citenamefont
  {Grason}(2014)}]{Azadi2014}%
  \BibitemOpen
  \bibfield  {author} {\bibinfo {author} {\bibfnamefont {A.}~\bibnamefont
  {Azadi}}\ and\ \bibinfo {author} {\bibfnamefont {G.~M.}\ \bibnamefont
  {Grason}},\ }\href {\doibase 10.1103/PhysRevLett.112.225502} {\bibfield
  {journal} {\bibinfo  {journal} {Physical Review Letters}\ }\textbf {\bibinfo
  {volume} {112}},\ \bibinfo {pages} {225502} (\bibinfo {year}
  {2014})}\BibitemShut {NoStop}%
\bibitem [{\citenamefont {Yao}(2017)}]{Yao2017}%
  \BibitemOpen
  \bibfield  {author} {\bibinfo {author} {\bibfnamefont {Z.}~\bibnamefont
  {Yao}},\ }\href {\doibase 10.1039/C7SM01599B} {\bibfield  {journal} {\bibinfo
   {journal} {Soft Matter}\ }\textbf {\bibinfo {volume} {13}},\ \bibinfo
  {pages} {5905} (\bibinfo {year} {2017})}\BibitemShut {NoStop}%
\bibitem [{\citenamefont {Jones}\ and\ \citenamefont {Sc}(1924)}]{LJ1924}%
  \BibitemOpen
  \bibfield  {author} {\bibinfo {author} {\bibfnamefont {J.~E.}\ \bibnamefont
  {Jones}}\ and\ \bibinfo {author} {\bibfnamefont {D.}~\bibnamefont {Sc}},\
  }\href {\doibase 10.1098/rspa.1924.0082} {\bibfield  {journal} {\bibinfo
  {journal} {Proc. R. Soc. Lond. A}\ }\textbf {\bibinfo {volume} {106}},\
  \bibinfo {pages} {463} (\bibinfo {year} {1924})}\BibitemShut {NoStop}%
\bibitem [{\citenamefont {Weeks}\ \emph {et~al.}(1971)\citenamefont {Weeks},
  \citenamefont {Chandler},\ and\ \citenamefont {Andersen}}]{WCA1971}%
  \BibitemOpen
  \bibfield  {author} {\bibinfo {author} {\bibfnamefont {J.~D.}\ \bibnamefont
  {Weeks}}, \bibinfo {author} {\bibfnamefont {D.}~\bibnamefont {Chandler}}, \
  and\ \bibinfo {author} {\bibfnamefont {H.~C.}\ \bibnamefont {Andersen}},\
  }\href {\doibase 10.1063/1.1674820} {\bibfield  {journal} {\bibinfo
  {journal} {The Journal of Chemical Physics}\ }\textbf {\bibinfo {volume}
  {54}},\ \bibinfo {pages} {5237} (\bibinfo {year} {1971})}\BibitemShut
  {NoStop}%
\bibitem [{\citenamefont {Stukowski}(2010)}]{OVITO}%
  \BibitemOpen
  \bibfield  {author} {\bibinfo {author} {\bibfnamefont {A.}~\bibnamefont
  {Stukowski}},\ }\href {\doibase 10.1088/0965-0393/18/1/015012} {\bibfield
  {journal} {\bibinfo  {journal} {Modelling and Simulation in Materials Science
  and Engineering}\ }\textbf {\bibinfo {volume} {18}},\ \bibinfo {pages}
  {015012} (\bibinfo {year} {2010})}\BibitemShut {NoStop}%
\bibitem [{\citenamefont {Stukowski}\ and\ \citenamefont
  {Arsenlis}(2012)}]{Stukowski2012}%
  \BibitemOpen
  \bibfield  {author} {\bibinfo {author} {\bibfnamefont {A.}~\bibnamefont
  {Stukowski}}\ and\ \bibinfo {author} {\bibfnamefont {A.}~\bibnamefont
  {Arsenlis}},\ }\href {\doibase 10.1088/0965-0393/20/3/035012} {\bibfield
  {journal} {\bibinfo  {journal} {Modelling and Simulation in Materials Science
  and Engineering}\ }\textbf {\bibinfo {volume} {20}},\ \bibinfo {pages}
  {035012} (\bibinfo {year} {2012})}\BibitemShut {NoStop}%
\bibitem [{\citenamefont {Pronk}\ and\ \citenamefont
  {Frenkel}(1999)}]{Pronk1999}%
  \BibitemOpen
  \bibfield  {author} {\bibinfo {author} {\bibfnamefont {S.}~\bibnamefont
  {Pronk}}\ and\ \bibinfo {author} {\bibfnamefont {D.}~\bibnamefont
  {Frenkel}},\ }\href {\doibase 10.1063/1.478339} {\bibfield  {journal}
  {\bibinfo  {journal} {The Journal of Chemical Physics}\ }\textbf {\bibinfo
  {volume} {110}},\ \bibinfo {pages} {4589} (\bibinfo {year}
  {1999})}\BibitemShut {NoStop}%
\bibitem [{Sup()}]{SupplementalMaterial}%
  \BibitemOpen
  \href@noop {} {}\bibinfo {howpublished} {{See Supplemental Material at [URL]
  for details on the dislocation core radius estimate and the additional strain
  components $\eta_{yy}$, $\eta_{xy}$, and $\eta_{yz}$.}}\BibitemShut {Stop}%
\bibitem [{\citenamefont {Mura}(1987)}]{Mura}%
  \BibitemOpen
  \bibfield  {author} {\bibinfo {author} {\bibfnamefont {T.}~\bibnamefont
  {Mura}},\ }in\ \href
  {http://link.springer.com/chapter/10.1007/978-94-009-3489-4_1} {\emph
  {\bibinfo {booktitle} {Micromechanics of defects in solids}}},\ \bibinfo
  {series and number} {\bibinfo {series} {Mechanics of {Elastic} and
  {Inelastic} {Solids}}\ No.~\bibinfo {number} {3}}\ (\bibinfo  {publisher}
  {Springer Netherlands},\ \bibinfo {year} {1987})\ pp.\ \bibinfo {pages}
  {1--73}\BibitemShut {NoStop}%
\bibitem [{\citenamefont {Chrzan}\ \emph {et~al.}(2010)\citenamefont {Chrzan},
  \citenamefont {Sherburne}, \citenamefont {Hanlumyuang}, \citenamefont {Li},\
  and\ \citenamefont {Morris}}]{Chrzan2010}%
  \BibitemOpen
  \bibfield  {author} {\bibinfo {author} {\bibfnamefont {D.~C.}\ \bibnamefont
  {Chrzan}}, \bibinfo {author} {\bibfnamefont {M.~P.}\ \bibnamefont
  {Sherburne}}, \bibinfo {author} {\bibfnamefont {Y.}~\bibnamefont
  {Hanlumyuang}}, \bibinfo {author} {\bibfnamefont {T.}~\bibnamefont {Li}}, \
  and\ \bibinfo {author} {\bibfnamefont {J.~W.}\ \bibnamefont {Morris}},\
  }\href {\doibase 10.1103/PhysRevB.82.184202} {\bibfield  {journal} {\bibinfo
  {journal} {Physical Review B}\ }\textbf {\bibinfo {volume} {82}},\ \bibinfo
  {pages} {184202} (\bibinfo {year} {2010})}\BibitemShut {NoStop}%
\bibitem [{\citenamefont {Anderson}\ \emph {et~al.}(2008)\citenamefont
  {Anderson}, \citenamefont {Lorenz},\ and\ \citenamefont
  {Travesset}}]{Anderson2008}%
  \BibitemOpen
  \bibfield  {author} {\bibinfo {author} {\bibfnamefont {J.~A.}\ \bibnamefont
  {Anderson}}, \bibinfo {author} {\bibfnamefont {C.~D.}\ \bibnamefont
  {Lorenz}}, \ and\ \bibinfo {author} {\bibfnamefont {A.}~\bibnamefont
  {Travesset}},\ }\href {\doibase 10.1016/j.jcp.2008.01.047} {\bibfield
  {journal} {\bibinfo  {journal} {Journal of Computational Physics}\ }\textbf
  {\bibinfo {volume} {227}},\ \bibinfo {pages} {5342} (\bibinfo {year}
  {2008})}\BibitemShut {NoStop}%
\bibitem [{\citenamefont {Glaser}\ \emph {et~al.}(2015)\citenamefont {Glaser},
  \citenamefont {Nguyen}, \citenamefont {Anderson}, \citenamefont {Lui},
  \citenamefont {Spiga}, \citenamefont {Millan}, \citenamefont {Morse},\ and\
  \citenamefont {Glotzer}}]{Glaser2015}%
  \BibitemOpen
  \bibfield  {author} {\bibinfo {author} {\bibfnamefont {J.}~\bibnamefont
  {Glaser}}, \bibinfo {author} {\bibfnamefont {T.~D.}\ \bibnamefont {Nguyen}},
  \bibinfo {author} {\bibfnamefont {J.~A.}\ \bibnamefont {Anderson}}, \bibinfo
  {author} {\bibfnamefont {P.}~\bibnamefont {Lui}}, \bibinfo {author}
  {\bibfnamefont {F.}~\bibnamefont {Spiga}}, \bibinfo {author} {\bibfnamefont
  {J.~A.}\ \bibnamefont {Millan}}, \bibinfo {author} {\bibfnamefont {D.~C.}\
  \bibnamefont {Morse}}, \ and\ \bibinfo {author} {\bibfnamefont {S.~C.}\
  \bibnamefont {Glotzer}},\ }\href {\doibase 10.1016/j.cpc.2015.02.028}
  {\bibfield  {journal} {\bibinfo  {journal} {Computer Physics Communications}\
  }\textbf {\bibinfo {volume} {192}},\ \bibinfo {pages} {97} (\bibinfo {year}
  {2015})}\BibitemShut {NoStop}%
\bibitem [{\citenamefont {Martyna}\ \emph {et~al.}(1996)\citenamefont
  {Martyna}, \citenamefont {Tuckerman}, \citenamefont {Tobias},\ and\
  \citenamefont {Klein}}]{Martyna1996}%
  \BibitemOpen
  \bibfield  {author} {\bibinfo {author} {\bibfnamefont {G.~J.}\ \bibnamefont
  {Martyna}}, \bibinfo {author} {\bibfnamefont {M.~E.}\ \bibnamefont
  {Tuckerman}}, \bibinfo {author} {\bibfnamefont {D.~J.}\ \bibnamefont
  {Tobias}}, \ and\ \bibinfo {author} {\bibfnamefont {M.~L.}\ \bibnamefont
  {Klein}},\ }\href {\doibase 10.1080/00268979600100761} {\bibfield  {journal}
  {\bibinfo  {journal} {Molecular Physics}\ }\textbf {\bibinfo {volume} {87}},\
  \bibinfo {pages} {1117} (\bibinfo {year} {1996})}\BibitemShut {NoStop}%
\bibitem [{\citenamefont {Chiu}(1977)}]{Chiu1977}%
  \BibitemOpen
  \bibfield  {author} {\bibinfo {author} {\bibfnamefont {Y.~P.}\ \bibnamefont
  {Chiu}},\ }\href {\doibase 10.1115/1.3424140} {\bibfield  {journal} {\bibinfo
   {journal} {Journal of Applied Mechanics}\ }\textbf {\bibinfo {volume}
  {44}},\ \bibinfo {pages} {587} (\bibinfo {year} {1977})}\BibitemShut
  {NoStop}%
\bibitem [{\citenamefont {Daw}(2006)}]{Daw2006}%
  \BibitemOpen
  \bibfield  {author} {\bibinfo {author} {\bibfnamefont {M.~S.}\ \bibnamefont
  {Daw}},\ }\href {\doibase 10.1016/j.commatsci.2006.02.009} {\bibfield
  {journal} {\bibinfo  {journal} {Computational Materials Science}\ }\textbf
  {\bibinfo {volume} {38}},\ \bibinfo {pages} {293} (\bibinfo {year}
  {2006})}\BibitemShut {NoStop}%
\bibitem [{\citenamefont {Gusev}\ \emph {et~al.}(1996)\citenamefont {Gusev},
  \citenamefont {Zehnder},\ and\ \citenamefont {Suter}}]{Gusev1996}%
  \BibitemOpen
  \bibfield  {author} {\bibinfo {author} {\bibfnamefont {A.~A.}\ \bibnamefont
  {Gusev}}, \bibinfo {author} {\bibfnamefont {M.~M.}\ \bibnamefont {Zehnder}},
  \ and\ \bibinfo {author} {\bibfnamefont {U.~W.}\ \bibnamefont {Suter}},\
  }\href {\doibase 10.1103/PhysRevB.54.1} {\bibfield  {journal} {\bibinfo
  {journal} {Phys. Rev. B}\ }\textbf {\bibinfo {volume} {54}},\ \bibinfo
  {pages} {1} (\bibinfo {year} {1996})}\BibitemShut {NoStop}%
\bibitem [{\citenamefont {Efron}\ and\ \citenamefont
  {Tibshirani}(1994)}]{Efron1994}%
  \BibitemOpen
  \bibfield  {author} {\bibinfo {author} {\bibfnamefont {B.}~\bibnamefont
  {Efron}}\ and\ \bibinfo {author} {\bibfnamefont {R.~J.}\ \bibnamefont
  {Tibshirani}},\ }\href@noop {} {\emph {\bibinfo {title} {An {Introduction} to
  the {Bootstrap}}}}\ (\bibinfo  {publisher} {CRC Press},\ \bibinfo {year}
  {1994})\ \bibinfo {note} {google-Books-ID: gLlpIUxRntoC}\BibitemShut
  {NoStop}%
\end{thebibliography}
\end{document}